 \documentclass[final,1p]{elsarticle}
\usepackage{amssymb}
\biboptions{sort&compress}
\usepackage{amsthm}
\usepackage{bm}
\usepackage{amsmath}
\usepackage{enumerate}
\usepackage{enumitem}
\DeclareMathOperator{\sech}{sech}
 \usepackage{lineno}
\usepackage[caption=false]{subfig}

\journal{}

\begin{document}

\begin{frontmatter}



\title{On the full dispersion Kadomtsev–Petviashvili equations for dispersive elastic waves}


\author[label1]{H. A. Erbay\corref{cor1}}
    \ead{husnuata.erbay@ozyegin.edu.tr}
    \address[label1]{Department of Natural and Mathematical Sciences, Faculty of Engineering, Ozyegin University,  Cekmekoy 34794, Istanbul, Turkey}
    \cortext[cor1]{Corresponding author}

\author[label1]{S. Erbay}
    \ead{saadet.erbay@ozyegin.edu.tr}

\author[label2]{A. Erkip}
    \ead{albert@sabanciuniv.edu}
    \address[label2]{Faculty of Engineering and Natural Sciences, Sabanci University, Tuzla 34956, Istanbul,  Turkey}

\begin{abstract}
Full dispersive models of water waves, such as the Whitham equation  and the full dispersion  Kadomtsev–Petviashvili (KP) equation,   are interesting from both the physical and mathematical points of view.  This paper studies analogous full dispersive KP models of nonlinear  elastic waves propagating in a nonlocal elastic medium. In particular we consider  anti-plane shear elastic waves which are assumed to be small-amplitude long waves. We propose two different full dispersive extensions of the  KP equation in the case of cubic nonlinearity and "negative dispersion". One of them is called the  Whitham-type  full dispersion  KP equation and the other one is called the BBM-type full dispersion KP equation. Most of the existing KP-type equations in the literature are particular cases of our full dispersion KP equations. We also introduce the simplified models of the new proposed full dispersion KP equations by approximating the operators in the equations.  We show that the line solitary wave solution of a simplifed form of the Whitham-type  full dispersion  KP equation is linearly unstable to long-wavelength transverse disturbances if the propagation speed of the line solitary wave is greater than a certain value. A similar analysis for a simplifed form of the BBM-type  full dispersion  KP equation does not provide a linear instability assessment.
\end{abstract}


\end{frontmatter}


\setcounter{equation}{0}
\section{Introduction}\label{sec1}

The  Kadomtsev–Petviashvili (KP) equation describes the propagation of weakly dispersive and small amplitude long waves propagating predominantly in one direction under weak transverse effects. It was proposed  by Kadomtsev and Petviashvilli  \cite{Kadomtsev1970}  to study  the stability of the one-soliton solution of the Korteweg-de-Vries (KdV) equation under the influence of weak transverse perturbations.  The KP equation with quadratic nonlinearity has been derived as a model equation in many different applications including surface and internal water waves,  ion-acoustic waves in plasmas, sound waves, nonlinear optics, lattice waves,  and dispersive elastic waves. Nonlinear generalizations of the KP equation with general power nonlinearities have also been extensively studied.

While the KP equation is widely used as an asymptotic model in the literature, it has been pointed out that it has certain shortcomings. In \cite{Lannes2003} it was shown that the KP equation is a poor asymptotic model at least for water waves. Another closely related issue is that the linear dispersion relation corresponding to the KP equation exhibits a singularity (the frequency  is not finite  as the wavenumber associated with the longitudinal direction tends to zero) but the parent equations of the original physical problem (for instance, the full water wave system) has no such  singularity. To eliminate the singularity we  require the absence of the Fourier components whose longitudinal wavenumbers are zero. But this imposes an artificial  zero-mass constraint  on the solutions of the KP equation.   In \cite{Lannes2013}  a fully dispersive version of the KP equation  has been proposed  to circumvent the difficulties in the KP equation. The basic idea is to replace the differential operator corresponding to the linear dispersion relation of the KP equation  by the pseudo-differential operator whose symbol corresponds to the exact dispersion relation of water waves. This is similar to how the Whitham equation is a fully dispersive version of the KdV equation \cite{whitham1974}. The full dispersion KP model has been studied from a number of viewpoints; see for instance \cite{Saut2013,Pilod2021,Klein2018} and the references therein.  The main conclusion to be drawn from those studies is that  the full dispersion KP equation  provides a more accurate asymptotic description of the weakly transverse waves than the KP equation and it imposes a weaker constraint than  the one for the KP equation.

In the present work our aim is to develop similar full-dispersion KP models for dispersive elastic waves, following the  approach proposed in \cite{whitham1974,Lannes2013} for water waves. We consider here anti-plane shear (displacement or strain) waves propagating  in a nonlocal elastic medium.   The nonlinear nonlocal wave equation derived previously in \cite{Erbay2011}  to describe the propagation of anti-plane shear waves in a nonlocal elastic medium  forms the basis of our subsequent discussion.  We refer the reader to \cite{Duruk2010,Eringen2002} for details concerning the nonlocal elasticity theory and to \cite{Horgan1995} for a general review about anti-plane shear deformations in solid mechanics.

We  first derive the  KP equation  with cubic nonlinearity  by using an asymptotic expansion technique. It is worth remarking that, with the plus sign in front of the transverse term, our derived equation is associated with the so-called KP II  equation (the "negative dispersion" case).  Then we propose two different forms of the full dispersion cubic KP equation: one is called the Whitham-type KP equation and the other is called the BBM-type KP equation. Both types of the full dispersion KP equations have the same linear dispersion relation,  but their nonlinearities are different, one local and one nonlocal nonlinear term. Some simplified forms of the full dispersion cubic KP equations are also presented and discussed in terms of how the simplified models are related to  the well known equations of dispersive wave propagation. A classic problem related to the KP-type equations is how small transverse perturbations affect the transverse stability of the unidirectional solitary waves.   In the last part of the present work, we  investigate linear instability of the line solitary wave solutions of the two particular forms of the Whitham-type  and  BBM-type KP equations  under long wavelength transverse perturbations.  We  show the existence of the transverse instability for the  Whitham-type KP equation if the propagation speed of the solitary wave is greater than a certain value. However, at the same order of approximation, a similar result is not obtained for the particular form of the BBM-type KP equation.

The content of the work is as follows. In the next section we introduce the nonlinear nonlocal wave equation describing the propagation of anti-plane shear waves in a nonlocal elastic medium. In Section \ref{sec3},   we derive the cubic KP equation   as the leading order approximation of  an asymptotic expansion for the elastic shear (displacement or strain) waves.  In Section \ref{sec4} we suggest the   Whitham-type  full dispersion  KP equation  for the anti-plane shear waves. In the same section  we also investigate linear instability of the line solitary waves for a simplified form of the Whitham-type  full dispersion  KP equation.  In Section \ref{sec5} we introduce the BBM-type  full dispersion  KP equation for  the anti-plane shear waves and again  investigate  linear instability of the line solitary waves for a simplified form of the BBM-type  full dispersion  KP equation.

\setcounter{equation}{0}
\section{A nonlocal equation governing anti-plane shear motions}\label{sec2}
We consider anti-plane shear motions in a homogeneous, isotropic, nonlinearly and nonlocally elastic medium. Let $x,y,z$ be rectangular Cartesian coordinates. Let $u_{1}, u_{2}, u_{3}$ be  the components of the displacement vector field at time $t$.   The anti-plane shear motion is defined by $u_{1}\equiv u_{2}\equiv 0$, $u_{3}=w(x,y,t)$, where $w$ is the out-of-plane displacement at position $(x,y)$ and time $t$. In  the absence of body forces, the differential equations of motion for the anti-plane shear motion reduce to a single scalar equation in $w$. The following equation expressed in terms of non-dimensional quantities has been suggested in \cite{Erbay2011} as a model equation:
\begin{equation}
        w_{tt}=\left( \beta \ast \frac{\partial F}{\partial w_{x}} \right)_x +\left( \beta \ast \frac{\partial F}{\partial w_{y}} \right)_y.
           \label{cau1}
\end{equation}
 Here  $F$ denotes the strain energy density function   and  the subscripts denote partial derivatives. A  pure anti-plane shear wave exists for some  special forms of the strain energy density function for isotropic elastic solids (we refer the reader to \cite{Destrade2019} and the references therein for  further details on this matter). As in \cite{Erbay2011}, here we assume that $F$  is a nonlinear function of $\vert \nabla w\vert^{2} \equiv (w_x^2+w_y^2)$  with $F(0)=0$.  The terms with $\beta$ in (\ref{cau1}) incorporate the nonlocal effects, where
\begin{equation*}
    (\beta \ast g)(x,y)=\iint_{\mathbb{R}^2}\beta (x-x^{\prime}, y-y^{\prime}) g(x^{\prime},y^{\prime}) ~dx^{\prime} dy^{\prime}
\end{equation*}
denotes convolution  of $\beta$ and $g$.   The kernel function $\beta(x,y)$ satisfies the normalization condition
\begin{equation}
  \iint_{\mathbb{R}^2}\beta(x,y) ~dxdy=1,  \label{normalization}
\end{equation}
and is a function of the modulus $\vert (x,y)\vert$, that is, $\beta(x,y)=\beta_{0}(\sqrt{x^2+y^2})$. We assume that $\beta$ and its Fourier transform   $\widehat{\beta}$ are sufficiently smooth functions.  When $\beta$ is taken as  the Dirac measure, (\ref{cau1}) reduces to the quasilinear wave equation for anti-plane shear motions in the context of the local (classical) theory of elasticity. So we may think of (\ref{cau1}) as a dispersive regularization of the classical wave equation for anti-plane shear motions. We refer the reader to \cite{Erbay2011} for some special examples of $\beta$ together with the corresponding nonlinear dispersive wave equations.

Henceforth, to simplify the computation,  we will assume that $F$ has the form
\begin{equation}
        F(S)=\frac{1}{2}S^2+\frac{\mu}{6}S^4, ~~~~~S=\vert \nabla w\vert=\sqrt{w_x^2+w_y^2}, \label{Fdef}
\end{equation}
where $\mu>0$ is a nonlinear elastic constant. With this choice of $F$, (\ref{cau1}) reduces to
\begin{equation}
         w_{tt}=\Big[ \beta \ast  \big(w_{x}[1+\frac{2}{3} \mu (w_x^2+w_y^2)]\big)\Big]_x+\Big[ \beta \ast  \big(w_{y}[1+\frac{2}{3} \mu (w_x^2+w_y^2)]\big)\Big]_y
           \label{cau2}
\end{equation}
or
\begin{equation}
         w_{tt}=\beta \ast  \Big[w_{xx}+w_{yy}+\frac{2}{3} \mu \Big( (3w_x^2+w_y^2)w_{xx}+4w_x w_y w_{xy}+(w_x^2+3w_{y}^2)w_{yy}\Big)\Big].
           \label{cau2s}
\end{equation}
The linear dispersion relation
\begin{equation}
        \omega^{2}=(k^2+l^2)\widehat{\beta}(k,l)   \label{disp}
\end{equation}
is obtained by neglecting  the cubic terms in (\ref{cau2s}), and then seeking solutions proportional to Fourier components, where $k$ and $l$  are the
wavenumbers (Fourier variables) corresponding to $x$ and $y$, respectively, $\omega$ is the frequency and $\widehat{\beta}(k,l)$ is the two-dimensional Fourier transform of $\beta(x,y) $. Notice that we have  $\widehat{\beta}(0,0)=1$ due to the normalization condition (\ref{normalization}). Furhermore, since $\beta(x,y)=\beta_{0}(\sqrt{x^2+y^2})$ is a radial function of space, the  Fourier transform of $\beta$ will be a radial function of Fourier variables, namely, a function of  $\sqrt{k^2+l^2}$.

We now introduce the operator $M$ with the symbol $m$ by
\begin{equation}
    M\big(D_{x}, D_{y}\big)g(x,y)={\cal F}^{-1}\Big[m(k,l){\cal F}g(x,y)\Big], ~~~~~    m(k,l)=\frac{1}{\widehat{\beta}(k,l)}-1,
\end{equation}
where $ {\cal F}$ and ${\cal F}^{-1}$ represent the two-dimensional Fourier transform and its inverse, respectively. Turning our attention to the fact that  $\widehat{\beta}(k,l)$  is a function of  $\sqrt{k^2+l^2}$, we notice that $M(D_{x}, D_{y})$ depends on $\sqrt{-\Delta}$, where  $\Delta$ is the two-dimensional Laplace operator $\Delta=D_{x}^2+D_{y}^2$. We have  $m(0,0)=0$ obviously due to $\widehat{\beta}(0,0)=1$. Using the operator $M$, we can rewrite  ( \ref{cau2s}) in the form
\begin{equation}
 \big[1+M\big(D_{x}, D_{y}\big)\big]w_{tt}=w_{xx}+w_{yy}+\frac{2}{3} \mu  \Big[ (3w_x^2+w_y^2)w_{xx}+4w_x w_y w_{xy}+(w_x^2+3w_{y}^2)w_{yy}\Big]
           \label{cau3}
\end{equation}
for which the linear dispersion relation is  $[1+m(k,l)]\omega^{2}=k^2+l^2$ and this  is just another way of expressing (\ref{disp}).

\setcounter{equation}{0}
\section{Derivation of the cubic KP equation}\label{sec3}

In this section our goal is to derive an evolution equation describing the propagation of small amplitude long waves by assuming that the waves mainly propagate along the positive $x$ direction  with weak  effect in the $y$ direction.    So the scales in the $x$ and $y$ directions are different and  the wavenumbers $k$ and $l$     along the longitudinal direction $x$ and  the transverse direction $y$, respectively, have different scales.  To reflect  the long-wave and weak transverse effect assumptions, we assume that $\vert k\vert \ll 1$ and $\vert l/k\vert \ll 1$. We note that, for long waves   with weak transverse effects,  the function $\widehat{\beta}(k,l)$ admits the expansion
\begin{equation}
    \widehat{\beta}(k,l)=1+\frac{1}{2}\frac{\partial^2 \widehat{\beta}}{\partial k^{2}}(0,0)k^2+\cdots,
\end{equation}
where we have used the normalization condition $\widehat{\beta}(0,0)=1$ and the fact that  $\widehat{\beta}(k,l)$ is a radial function in Fourier space, that is, $\frac{\partial\widehat{\beta}}{\partial k}(0,0)=\frac{\partial\widehat{\beta}}{\partial l}(0,0)=0$. So the function $m(k,l)$ admits the expansion $m(k,l)= 2\nu k^{2}+\cdots$, where we introduce the parameter $\nu=-\frac{1}{4}\frac{\partial^2 \widehat{\beta}}{\partial k^{2}}(0,0)$. We assume that the parameter $\nu$ is positive, which is suggested by the kernel functions  that are often used in the literature \cite{Erbay2011}.   Then, taking the positive square root in (\ref{disp}) we get the dispersion relation for the right-going  waves in  the form
\begin{equation}
        \omega= k \big(1+\frac{1}{2}\frac{l^2}{k^2}+\cdots\big) \big(1-\nu k^2+\cdots\big)=k \big(1-\nu k^2+\frac{1}{2}\frac{l^2}{k^2}+\cdots\big).  \label{dis-app}
\end{equation}
Notice that the phase of the long waves with weak transverse effects can be approximated as $kx+ly-\omega t = k(x-t)+ly+k(\nu k^{2}-\frac{1}{2}\frac{l^2}{k^2})t+\cdots$. So we introduce the scaled variables
\begin{equation}
        X=\varepsilon (x-t), ~~~~Y=\varepsilon^2 y, ~~~~ T=\varepsilon^{3}t
\end{equation}
to describe the right-going long waves with weak transverse perturbation, where $\varepsilon$ is a small parameter characterizing the smallness of the wavenumber in the $x$ direction, with a much smaller wavenumber in the transverse direction. Using the notation $w(x,y,t)=w(X,Y,T)$, we rewrite (\ref{cau3}) in terms of the scaled variables as
\begin{eqnarray}
&& \!\!\!\!\!\!\!\!\!\!\!\!\!\!\!\!\!\!\!
[1+M\big(\varepsilon D_{X}, \varepsilon^2 D_{Y}\big)]\big( w_{XX}-2\varepsilon^2 w_{XT}+\varepsilon^4 w_{TT} \big) \nonumber \\
&&~~~~~~~~~~~  = w_{XX}+\varepsilon^2 \big( w_{YY}+2 \mu  w_X^{2} w_{XX}\big)  \nonumber \\
  &&~~~~~~~~~~~~~~                              +\frac{2}{3} \varepsilon^4 \mu \big( w_Y^2w_{XX}+4 w_X w_Y w_{XY}+w_X^2w_{YY}\big)
                                +2  \varepsilon^6 \mu  w_Y^2w_{YY}.
           \label{cau4}
\end{eqnarray}
 Expanding $w(X,Y,T)$ to a Taylor series in $\varepsilon^2$ in the form $w(X,Y,T)=u(X,Y,T)+{\cal O}(\varepsilon^2)$, using the approximation  $M\big(\varepsilon D_{X}, \varepsilon^2 D_{Y}\big)=-2\nu\varepsilon^2 D_{X}^2+{\cal O}(\varepsilon^4)$, substituting the series expansions into (\ref{cau4}) and equating coefficients of like powers of $\varepsilon$ we get a hierarchy of equations. At the leading order we get the cubic KP equation
\begin{equation}
        u_{XT}+ \mu (u_{X})^{2}u_{XX}+\nu u_{XXXX}+\frac{1}{2}u_{YY}=0, \label{mKP}
\end{equation}
which describes a balance among  weak nonlinearity, weak dispersion and transverse effect. At this point we remind that the classical KP equation appears with quadratic nonlinearity \cite{Kadomtsev1970}. If we replaced the quartic nonlinearity assumption regarding $F$ in (\ref{Fdef})  by the cubic nonlinearity assumption, we would get the classical KP equation using a similar approach. The above cubic KP equation corresponding to the "negative dispersion" case appears as a physically relevant model in important physical applications such as   sound waves in antiferromagnetics \cite{Turitsyn1985},   elastic shear waves \cite{Destrade2011} and micropolar elastic waves \cite{Erbay1999}.

We close this part by stating (\ref{mKP}) in the original coordinates $x$, $y$ and $t$. Since $\varepsilon^4 D^2_{XT}=D_{xt}+D_{xx}$, $\varepsilon^4D_{X}^{4}=D_{x}^{4}$ and $\varepsilon^4 D_{Y}^{2}=D_{y}^{2}$,  the cubic KP equation
\begin{equation}
        w_{xt}+w_{xx}+ \mu (w_{x})^{2}w_{xx}+\nu w_{xxxx}+\frac{1}{2}w_{yy}=0 \label{mKP-or}
\end{equation}
is  a model equation for  anti-plane shear  waves under the assumptions small-amplitude, small wavenumber and  weak transverse effect. The linear dispersion relation of this equation coincides with   the dispersion relation (\ref{dis-app}) of the linear anti-plane shear waves exactly  if the higher-order terms in (\ref{dis-app}) are neglected. In order to state (\ref{mKP-or}) in terms of the strain component $v$ defined by $v=w_{x}$ rather than the displacement $w$, we differentiate it with respect to $x$ and make the substitution $v=w_{x}$:
\begin{equation}
        \big(v_{t}+v_{x}+\mu v^{2}v_{x}+\nu v_{xxx}\big)_{x}+\frac{1}{2}v_{yy}=0. \label{mKP-pot}
\end{equation}
When $y$-dependence is omitted, (\ref{mKP-pot}) reduces to the modified KdV equation. So the $\sech$-type solitary wave solutions of the modified KdV equation are  automatically planar solutions of (\ref{mKP-pot}). For the KP equation with quadratic nonlinearity,  a linear transverse instability analysis of the line solitary wave solution  has been performed in \cite{Pego1997} and no unstable modes have been found.

\setcounter{equation}{0}
\section{The Whitham-type  full dispersion  KP equation}\label{sec4}

One drawback of the KP equation is the existence of  the singularity of the linear dispersion relation (\ref{dis-app})  at $k=0$.  Since this is not present in the original dispersion relation (\ref{disp}), it makes the KP equation  a poor approximation to the original system. In order to remedy such shortcomings of the KP equation, a full dispersion KP equation for water waves was proposed  in \cite{Lannes2013}. We refer the reader to \cite{Saut2013,Pilod2021,Klein2018} for various properties of the full dispersion KP equation. The most important aspect of the full dispersion KP equation is that  the linear dispersion relation  is the same as the exact dispersion relation of water waves model. This is the reason why  a larger region of validity for the full dispersion KP equation proposed in \cite{Lannes2013} is expected. In the present section and the next one we extend the idea of the full dispersion model for water waves to dispersive elastic waves and propose two different full dispersion KP models for the anti-plane shear waves discussed in the preceding sections. In the present section we will concentrate on how the cubic KP equation derived in the previous section can be replaced by a Whitham-type  full dispersion  KP equation  and then will perform a linear transverse instability analysis of the line solitary wave for a particular case.

Using the previous approach outlined in  \cite{Lannes2013}, we replace the linear dispersion relation for (\ref{mKP-or}) by the exact dispersion relation $\omega=k\sqrt{1+l^2/k^2}\:/\sqrt{1+m(k,l)}$.  Then, for anti-plane shear waves propagating in nonlocal elastic media  we get the Whitham-type full dispersion  KP equation as
\begin{equation}
        w_{xt}+L(D_{x},D_{y})w_{xx}+ \mu (w_{x})^{2}w_{xx}=0 \label{fmKP-or}
\end{equation}
in terms of the displacement $w$, or as
\begin{equation}
        v_{t}+L(D_{x},D_{y})v_{x}+ \mu v^{2}v_{x}=0 \label{fmKP-str}
\end{equation}
in terms of  the strain component $v=w_{x}$. Here the nonlocal operator $L(D_{x},D_{y})$  takes into account the full dispersive  effect and  is defined by
\begin{equation}
        L(D_{x},D_{y})=\frac{\big(1+\frac{D_{y}^2}{D_{x}^2}\big)^{1/2}}{\big[1+M\big(D_{x}, D_{y}\big) \big]^{1/2}}. \label{L-op}
\end{equation}
One can readily see that the symbol $p(k,l)$ of $L(D_{x},D_{y})$ is given by
\begin{equation}
       p(k,l)=\frac{\big(1+\frac{l^2}{k^2}\big)^{1/2}}{\big[1+m(k,l) \big]^{1/2}}
                =       \big[(1+\frac{l^2}{k^2})\widehat{\beta}(k,l)\big]^{1/2}, \label{sym}
\end{equation}
for which $L\big(D_{x}, D_{y}\big)w(x,y)={\cal F}^{-1}\big[p(k,l){\cal F}w(x,y)\big]$. It is worth pointing out that the singularity at $k=0$ in the linear dispersion relation of (\ref{mKP-pot}) does not arise in the linear dispersion relation
\begin{equation}
        \omega= k p(k,l)=\frac{\big(k^2+l^2\big)^{1/2}}{\big[1+m(k,l) \big]^{1/2}}
                =       \big[(k^2+l^2)\widehat{\beta}(k,l)\big]^{1/2}
\end{equation}
corresponding to   (\ref{fmKP-str}).

The three conserved quantities for (\ref{fmKP-str}) are
\begin{equation}
       Q= \iint_{\mathbb{R}^2}v^{2} ~dxdy, ~~~~
       E= \iint_{\mathbb{R}^2}\frac{1}{2}\Big[(L^{1/2}v)^{2}+\frac{\mu}{6}v^{4}\Big] ~dxdy,~~~~
       P= \iint_{\mathbb{R}^2}v ~dxdy.  \label{conser}
\end{equation}
Notice that if we set
\begin{displaymath}
  1+M(D_{x},D_{y})=\frac{1+\frac{D_{y}^2}{D_{x}^2}}{\big(1+\frac{1}{2}\frac{D_{y}^2}{D_{x}^2}+\nu D_{x}^2\big)^2},
\end{displaymath}
the operator $L$ will be of the form $L(D_{x},D_{y})=1+\frac{1}{2}\frac{D_{y}^2}{D_{x}^2}+\nu D_{x}^2$. For this special form of $L$ the Whitham-type full dispersion KP equation (\ref{fmKP-str}) reduces to the cubic KP equation (\ref{mKP-pot}) (where we first differentiate (\ref{fmKP-str}) with respect to $x$ and then substitute $L$ into the resulting equation) and  the conserved quantity $E$ (energy) defined in (\ref{conser}) becomes
\begin{equation}
       E= \iint_{\mathbb{R}^2}\frac{1}{2}\Big[v^{2}-\nu (v_{x})^{2}+\frac{1}{2}\big(D_{x}^{-1}v_{y}\big)^{2}+\frac{\mu}{6}v^{4}\Big] ~dxdy. \label{conserKP}
\end{equation}
As another special case, if we neglect transverse effects  in (\ref{fmKP-str})(that is,  if we assume that the waves propagate exactly along the $x$ direction), it reduces to
\begin{equation}
    v_{t}+L(D_{x},0)v_{x}+ \mu v^{2}v_{x}=0,
\end{equation}
where $L(D_{x},0)= [1+M(D_{x},0)]^{-1/2}$ and the symbol of $L(D_{x},0)$ is $\big[\widehat{\beta}(k,0)\big]^{1/2}$. In other words, in the absence of transverse effects, we get the integro-differential equation
\begin{equation}
        v_{t}+\int_{\mathbb{R}}\gamma( x-x^{\prime})v_{x}(x^{\prime},t)dx^{\prime}+ \mu v^{2}v_{x}=0, \label{nmKdV}
\end{equation}
where the kernel function $\gamma$ is the inverse Fourier transform of $\big[\widehat{\beta}(k,0)\big]^{1/2}$ with respect to $k$ only. We observe that  (\ref{nmKdV}) is a Whitham-type wave equation \cite{whitham1974} and it becomes the modified Whitham equation for $\widehat{\beta}(k,0)=(\tanh k)/k$.  So the nonlocal KP equation (\ref{fmKP-str}) can be considered as  a natural two-dimensional version of the  Whitham-type wave equation (\ref{nmKdV}) when weak transverse perturbations are incorporated.  Furthermore,  if we take $\big[\widehat{\beta}(k,0)\big]^{1/2}=1/(1+k^2)$, then the kernel $\gamma$ in (\ref{nmKdV}) becomes $\gamma( x )=\frac{1}{2}e^{-\vert x\vert}$ which is  the Green’s function for the operator $1-D_{x}^2$. In such a case, the integro-differential equation (\ref{nmKdV}) reduces to the modified  Fornberg–Whitham equation \cite{Fornberg1978}:
\begin{equation}
         v_{t}+v_{x}-v_{xxt}+ \mu v^{2}v_{x}-\mu(v^{2}v_{x})_{xx}=0.  \label{fornW}
\end{equation}
The last special case we shall consider for (\ref{fmKP-str}) is the case of
\begin{displaymath}
  1+M(D_{x},D_{y})=\frac{(1-D_{x}^{2})^{2}(1+\frac{D_{y}^2}{D_{x}^2})}{\big(1+\frac{1}{2}\frac{D_{y}^2}{D_{x}^2}\big)^2}
\end{displaymath}
in which $L(D_{x},D_{y})=(1+\frac{1}{2}\frac{D_{y}^2}{D_{x}^2})/(1-D_{x}^{2})$. If we differentiate (\ref{fmKP-str}) with respect to $x$ and then substitute $L$ into the resulting equation, we get
\begin{equation}
         \Big(v_{t}+v_{x}-v_{xxt}+\mu v^{2}v_{x}-\mu(v^{2}v_{x})_{xx}\Big)_{x}+\frac{1}{2}v_{yy}=0.  \label{2DfornW}
\end{equation}
We remark that (\ref{2DfornW}) is the two-dimensional version of the  modified  Fornberg–Whitham equation (\ref{fornW}) when weak transverse effects are incorporated.

\subsection{Linear transverse instability for a simplified model}

In this subsection we study the linear transverse instability of the unidirectional solitary waves for a simplified form of the Whitham-type full dispersion equation (\ref{fmKP-str}). If the following expansions
\begin{equation}
        [1+M\big(D_{x}, D_{y}\big) \big]^{-1/2}\approx 1-\frac{1}{2}M\big(D_{x}, D_{y}\big) ~~~\text{and}~~~
        \big(1+\frac{D_{y}^2}{D_{x}^2}\big)^{1/2}\approx 1+\frac{1}{2}\frac{D_{y}^2}{D_{x}^2}    \label{op-exp}
\end{equation}
 are used,     (\ref{fmKP-str})  then  becomes
\begin{equation}
        v_{xt}+\big[1-\frac{1}{2}M\big(D_{x}, D_{y}\big)\big]\big(v_{xx}+\frac{1}{2}v_{yy}\big)+\frac{\mu}{3} (v^{3})_{xx}=0. \label{smKP-str}
\end{equation}
Furthermore if we neglect transverse effects  and eliminate one $x$ derivative in the resulting equation by assuming that $v$ and its derivatives vanish at infinity, (\ref{smKP-str}) becomes
\begin{equation}
        v_{t}+\big[1-\frac{1}{2}M\big(D_{x}, 0\big)\big]v_{x}+\frac{\mu}{3} (v^{3})_{x}=0. \label{smKP-uni}
\end{equation}
So, every solution of (\ref{smKP-uni}) is also a $y$-independent solution of (\ref{smKP-str}).

For the simplest case $M(D_{x}, D_{y})=-2\nu D_x^2$, (\ref{smKP-str}) becomes
\begin{equation}
        \Big(v_{t}+v_{x}+\frac{ \mu}{3} (v^{3})_{x}+\nu v_{xxx}\Big)_{x}+\frac{1}{2}(v_{yy}+\nu v_{xxyy})=0. \label{W-ins}
\end{equation}
Notice that in the absence of the term $v_{xxyy}$ this equation reduces to the cubic KP equation (\ref{mKP-pot}). Also, we remind that the unidirectional solitary wave solutions of the KP equation are linearly stable under weak transverse perturbations (We refer the reader to \cite{Pego1997} for the quadratic nonlinearity case and to \cite{Kataoka2004} for the general power nonlinearity case). It is of interest to explore how the additional term $v_{xxyy}$ affects linear transverse stability of the solitary waves. Thus, in the remainder of this subsection, we will investigate the linear transverse instability of the line solitary wave solutions of (\ref{W-ins}).

If we neglect transverse effects, (\ref{W-ins}) reduces to the modified KdV equation
\begin{equation}
        v_{t}+v_{x}+\mu  v^{2}v_{x}+\nu v_{xxx}=0. \label{mkdv}
\end{equation}
Consider a traveling wave solution of (\ref{mkdv}) in the form $v(x,t)=\alpha R(\xi)$ where  $\xi=x-ct$ is the travelling wave coordinate and $\alpha$ and $c$ are constants. It follows that $R(\xi)$ satisfies the ordinary differential equation
\begin{equation}
  {\cal N}R\equiv \Big[ \frac{d^2}{d \xi^2}  +\kappa^2 \big(2 R^2(\xi) -1\big)\Big]R=0 \label{Nop}
\end{equation}
with $\mu>0$, $\nu >0$, $c-1>0$ and
\begin{equation}
       \alpha^2=6(c-1)/\mu,~~~~\kappa^{2}=(c-1)/\nu. \label{param}
\end{equation}
Recalling that  we are interested in solutions of (\ref{mkdv}) which  decay  to zero at infinity, we note that a decaying solution of (\ref{Nop}) is $R(\xi)=\sech (\kappa\xi)$. Thus, the solitary wave solution of (\ref{mkdv}) is obtained as $v(x,t)=\alpha \sech \big(\kappa (x-ct)\big)$ in which the wave propagates with speed $c$ in the positive direction of the $x$-axis. This solution is  a $y$-independent solution of (\ref{W-ins}). We now consider small transverse perturbations of the unidirectional solitary wave solution of (\ref{W-ins}). Furthermore we assume that the perturbations are periodic in $y$ with long wavelength.  As a result, we look for solutions of (\ref{W-ins}) in the form
\begin{equation}
       v(x,y,t)=\alpha \sech (\kappa\xi) +z(\xi)e^{i\lambda y+\Omega t},  \label{pertur}
\end{equation}
where $\xi=x-ct$ is  the travelling wave coordinate, $\lambda$ is a real constant representing the transverse wave number and $\Omega$ is possibly a complex constant and the existence of the solitary wave requires $c>1$. We also assume that $z$ and its derivatives vanish in the limit $\xi \rightarrow \pm \infty$. Substituting (\ref{pertur}) into (\ref{W-ins}), linearizing the resulting equation about the solitary-wave solution $\alpha \sech (\kappa\xi)$ and using the relation $\alpha^{2}=6\nu \kappa^{2}/\mu $, after lengthy calculations  we get   the fourth-order linear ordinary differential equation
\begin{equation}
         \frac{d^2}{d \xi^2}{\cal L} z=\frac{1}{2} \lambda^2 \frac{d^2z}{d \xi^2}
  -\frac{\Omega}{\nu} \frac{d z}{d \xi}+ \frac{\lambda^2}{2\nu}   z, \label{ode}
\end{equation}
where the linear operator ${\cal L}$ is defined by
\begin{equation}
       {\cal L}=\frac{d^2}{d \xi^2}  +\kappa^2 \big( 6\sech^2(\kappa\xi) -1\big). \label{oper}
\end{equation}
The solitary wave solution $\alpha \sech (\kappa\xi)$ of (\ref{W-ins}) is transversely unstable if (\ref{ode})  has solutions for which  $\Omega$ has a positive real part.  Under the assumption of long-wavelength transverse perturbations, $\vert \lambda \vert <<1$, we approximate the solution by the small parameter expansion
\begin{equation}
        z(\xi)= z_0(\xi)+ z_1(\xi) \lambda +  z_2(\xi) \lambda^2+\cdots,~~~~~  \Omega = \Omega_1 \lambda+ \Omega_2 \lambda^2 +\cdots .
        \label{expan}
 \end{equation}
Substitution of  the series expansions into (\ref{ode})   leads to a hierarchy of equations if we equate coefficients of like powers of  $\lambda$. At the lowest order, we get  the homogeneous equation
\begin{equation}
    \mathcal{L} z_0=0    ~~\text{~or~}~~ \frac{d^2z_0}{d \xi^2}+\kappa^2 \big(6  \sech^2(\kappa\xi) -1\big) z_0=0. \label{zeroth}
 \end{equation}
 To get a solution of this equation we first observe that $\big({\cal N}R\big)^{\prime}= {\cal L}R^{\prime}=0$, which is  obtained by differentiating (\ref{Nop}) with repect to $\xi$ and using the definitions of the  operators $\cal N$ and $\cal L$ (hereafter the prime denotes the derivative with respect to $\xi$). Since $R(\xi)=\sech (\kappa \xi)$, a decaying solution of the homogeneous equation (\ref{zeroth}) is obtained as
\begin{displaymath}
    z_0(\xi)=-a_0\kappa\sech (\kappa\xi) \tanh (\kappa \xi) =a_0 (\sech (\kappa\xi))^{\prime},
\end{displaymath}
where $a_{0}$ is an arbitrary constant. To simplify the presentation, henceforth we will use the notation  $z_0(\xi)=a_0 R^{\prime}(\xi)$. At ${\cal O}(\lambda)$ we get from (\ref{ode}) the following inhomogeneous differential equation for $z_{1}$
\begin{equation}
   \frac{d^2}{d \xi^2}\mathcal{L} z_1
                                =-\frac{\Omega_1}{\nu}\frac{d z_0}{d\xi}. \label{firsta}
 \end{equation}
 If we substitute $z_0(\xi)=a_0 R^{\prime}(\xi)$ to the right-hand side, integrate twice and use the zero conditions at infinity, (\ref{firsta}) reduces to
 \begin{equation}
        \mathcal{L}z_1=-\frac{\Omega_1}{\nu} a_0 R(\xi). \label{firstb}
 \end{equation}
 A decaying solution of this inhomogeneous differential equation is
\begin{equation}
    z_1(\xi)= a_1R^{\prime}(\xi)-\frac{\Omega_1 a_0}{2\kappa^3\nu}\big(\kappa\xi-\coth (\kappa\xi)\big)R^{\prime}(\xi), \label{firsts}
 \end{equation}
 where $a_{1}$ is an arbitrary constant. Substituting (\ref{expan}) into (\ref{ode}), at ${\cal O}(\lambda^{2})$ we arrive at the inhomogeneous differential equation
 \begin{equation}
        \frac{d^2}{d \xi^2}\mathcal{L} z_2
                        = \frac{1}{2}\frac{d^{2}z_0}{d \xi^2} -\frac{1}{\nu}\big(\Omega_1\frac{d z_1}{d \xi} +\Omega_2\frac{d z_0}{d \xi}\big)  +\frac{1}{2\nu} z_0 \label{seconda}
 \end{equation}
 for $z_{2}$. To derive a solvability condition for this equation we  multiply both sides of (\ref{seconda}) by $r(\xi)$ for which with $r^{\prime}=R$  and then integrate over the entire real line. Using the self-adjointness of $\mathcal{L}$, we see that the integral obtained from  the left-hand side of  (\ref{seconda}) is identically zero:
 \begin{equation}
                \big\langle (\mathcal{L}z_2)^{\prime\prime},r\big\rangle  = \big\langle \mathcal{L}z_2,r^{\prime\prime}\big\rangle
                                        = \big\langle z_2, \mathcal{L}r^{\prime\prime}\big\rangle = \big\langle z_2, \mathcal{L}R^{\prime}\big\rangle=0,  \label{left}
 \end{equation}
 where  the inner product is defined as $\big\langle f(x), g(x) \big\rangle=\int_{\mathbb{R}}f(x)g(x)dx$.  One can show that the following relations hold:
  \begin{eqnarray*}
       &&   \big\langle R,R\big\rangle=\frac{2}{\kappa},~~~~\big\langle R^{\prime},R\big\rangle=0,~~~~\big\langle R^{\prime\prime},R\big\rangle=-\frac{2}{3}\kappa  \\
       &&  \big\langle \kappa\xi R^{\prime},R\big\rangle=-\frac{\kappa}{2}\big\langle R,R\big\rangle =-1,~~~~
       \big\langle R^{\prime}\coth (\kappa\xi),R\big\rangle=-\kappa\big\langle R,R\big\rangle=-2.
  \end{eqnarray*}
 Using these results, we obtain the values of the integrals associated with the terms on the right-hand side of (\ref{seconda}) as
 \begin{eqnarray}
       &&  \!\!\!\!\!\!\!\!\!\!\!
       \big\langle z_{0},r\big\rangle
                                 = a_{0}\big\langle R^{\prime},r\big\rangle=- a_{0}\big\langle R,r^{\prime}\big\rangle=-a_{0}\big\langle R,R\big\rangle=-\frac{2}{\kappa} a_{0}  \label{zzz1}\\
      && \!\!\!\!\!\!\!\!\!\!\!
      \big\langle z_{0}^{\prime},r\big\rangle
                                 = a_{0}\big\langle R^{\prime\prime},r\big\rangle =-a_{0}\big\langle R^{\prime},r^{\prime}\big\rangle
                                                =-a_{0}\big\langle R^{\prime},R\big\rangle= 0 \label{zzz2}\\
      &&   \!\!\!\!\!\!\!\!\!\!\!
      \big\langle z_{0}^{\prime\prime},r\big\rangle
                                 = a_{0}\big\langle R^{\prime\prime\prime},r\big\rangle
                                =a_{0}\big\langle R^{\prime},r^{\prime\prime}\big\rangle =
                                a_{0}\big\langle R^{\prime},R^{\prime}\big\rangle=\frac{2}{3}\kappa a_{0}
                                \label{zzz3}\\
      &&   \!\!\!\!\!\!\!\!\!\!\!
      \big\langle z_{1}^{\prime},r\big\rangle
                                =-\big\langle z_{1},r^{\prime}\big\rangle=-\big\langle z_{1},R\big\rangle                             \nonumber \\
      &&~~~                    =-a_1\big\langle R^{\prime},R\big\rangle +\frac{\Omega_1
                                a_0}{2\kappa^3\nu}\big\langle \kappa\xi R^{\prime},R\big\rangle
                               -\frac{\Omega_1
                                a_0}{2\kappa^3\nu}\big\langle R^{\prime} \coth (\kappa\xi),R\big\rangle \nonumber  \\
        &&~~~                         =\frac{\Omega_1 }{2\kappa^3\nu}a_0. \label{zzz4}
 \end{eqnarray}
 Combining these results with (\ref{left}) we get the solvability condition for (\ref{seconda}) in the form
 \begin{equation}
                0=\frac{1}{3}\kappa   a_{0}-\frac{\Omega_1^{2} }{2\kappa^3\nu^{2}}a_0 -\frac{1}{\kappa\nu}a_{0}     \label{solv1}
 \end{equation}
 or, with $a_{0}\neq 0$
 \begin{equation}
               \Omega_{1}^{2}=\frac{2}{3}\kappa^{2}\nu(\kappa^2\nu-3)=\frac{2}{3}(c-1)(c-4),     \label{solvf}
 \end{equation}
 where we have used the relation $\kappa^{2}=(c-1)/\nu$. Since $\Omega_{1}^{2}>0$ for $c>4$, the linear instability is determined by the magnitude of the wave speed $c$, or equivalently, by the magnitude of the amplitude $\alpha$. Recall that the  wave speed $c$ depends on the amplitude $\alpha$ through the relation  $\alpha^2=6(c-1)/\mu$.  Equation (\ref{solvf}) clearly shows that when $c>4$ (or when $\alpha^{2} > 18/\mu$), the  line solitary wave solutions of (\ref{W-ins}) are linearly unstable against the long-wavelength periodic transverse perturbations.   It is worth mentioning that this is in contrast to what was observed for the standard KP equation (that is, the one with the absence of the term $v_{xxyy}$).   A linear instability analysis has been performed  in \cite{Pego1997} and \cite{Kataoka2004}   for the standard KP equation with quadratic and general power nonlinearity, respectively, and no unstable modes have been found.

\setcounter{equation}{0}
\section{The BBM-type full dispersion KP equation}\label{sec5}

In the present section, inspired by the so-called BBM trick, we propose a different way to get a full dispersion KP equation for anti-plane shear waves propagating in nonlocal elastic media. Following again  \cite{Lannes2013}, we replace the linear dispersion relation  of (\ref{mKP-or}) by the exact dispersion relation $\sqrt{1+m(k,l)}\:\omega=k\sqrt{1+l^2/k^2}$.  However, contrary to the one considered in the previous section, we distribute the dispersive effect into the first two terms of (\ref{mKP-or}). Then, we reach the  full dispersion cubic KP equation
\begin{equation}
       \big[1+M\big(D_{x}, D_{y}\big) \big]^{1/2} w_{xt}+\big(1+\frac{D_{y}^2}{D_{x}^2}\big)^{1/2}w_{xx}+  \mu (w_{x})^{2}w_{xx}=0 \label{BmKP-or}
\end{equation}
in terms of the displacement $w$ for anti-plane shear waves. We underline that for the linear parts of (\ref{BmKP-or}) and (\ref{fmKP-or}) we have the same nonlocal operator   taking into account the full dispersive  effect.  In terms of  the strain component $v=w_{x}$,  (\ref{BmKP-or})   becomes
\begin{equation}
         \big[1+M\big(D_{x}, D_{y}\big) \big]^{1/2} v_{t}+\big(1+\frac{D_{y}^2}{D_{x}^2}\big)^{1/2}v_{x}+ \mu v^{2}v_{x}=0. \label{BmKP-str}
\end{equation}
The three conserved quantities for (\ref{BmKP-str}) are
\begin{eqnarray}
       &&Q= \iint_{\mathbb{R}^2}\Big[(1+M)^{1/4}v\Big]^2 ~dxdy, ~~~~
       E= \iint_{\mathbb{R}^2}\frac{1}{2}\Big[ \Big(\big(1+\frac{D_{y}^2}{D_{x}^2}\big)^{1/4}v\Big)^{2}+\frac{\mu}{6}v^{4}\Big] dxdy, \nonumber \\
       &&P= \iint_{\mathbb{R}^2}(1+M)^{1/2}v ~dxdy. \label{conserbbm}
\end{eqnarray}
As a special case, if we  assume that the waves propagate exactly along the $x$ direction, (\ref{BmKP-or}) reduces to
\begin{equation}
       \big[1+M\big(D_{x}, 0\big) \big]^{1/2} v_{t}+v_{x}+\mu  v^{2}v_{x}=0, \label{BmKdV}
\end{equation}
or to the integro-differential equation
\begin{equation}
        v_{t}+\int_{\mathbb{R}}\gamma( x-x^{\prime})\big[v_{x}(x^{\prime},t)+ \mu (v^{2}v_{x})(x^{\prime},t)\big]dx^{\prime}=0, \label{BBMint}
\end{equation}
where the kernel function $\gamma$ is the inverse Fourier transform of $\big[\widehat{\beta}(k,0)\big]^{1/2}$. We observe that if we take $\big[\widehat{\beta}(k,0)\big]^{1/2}=1/(1+k^2)$ in which $\big[1+M\big(D_{x}, 0\big) \big]^{1/2}=1-D_{x}^{2}$, (\ref{BBMint}) reduces to the  cubic BBM equation \cite{Benjamin1972}, which is the reason why we may call (\ref{BmKdV}) as the BBM-type nonlocal  equation. Equation  (\ref{BmKP-or}) or (\ref{BmKP-str}) (hereafter called the BBM-type full dispersion KP equation) can be considered as  a natural two-dimensional version of  (\ref{BmKdV}) when weak transverse perturbations are incorporated. The most important difference of the full dispersion KP models (\ref{fmKP-or}) and (\ref{BmKP-or}) is that (\ref{BmKP-or}) contains a nonlocal nonlinear term while (\ref{fmKP-or}) contains a local nonlinear term (a comparison between the special cases (\ref{BBMint}) and (\ref{nmKdV}) can help us to see this easily).

\subsection{Linear transverse instability for a simplified model}
This subsection is devoted to  a linear transverse instability analysis of the unidirectional solitary waves for a simplified form of the BBM-type full dispersion equation (\ref{BmKP-str}). If the  expansions given  in (\ref{op-exp})  are used,    (\ref{BmKP-str})   reduces to
\begin{equation}
       \big[1+\frac{1}{2}M\big(D_{x}, D_{y}\big)\big] v_{xt}+v_{xx}+\frac{1}{2}v_{yy}+\frac{\mu}{3} (v^{3})_{xx}=0. \label{sBKP-str}
\end{equation}
For the particular case $M=M\big(D_{x}\big)$, similar equations are available in the literature. For instance, the reader is referred to equation (1.1) of \cite{Bona2002} (general power nonlinearity case) and to  equation (11) of \cite{Klein2012} (quadratic nonlinearity case). If we neglect transverse effects  and eliminate one $x$ derivative in the resulting equation under the zero conditions at infinity,  (\ref{sBKP-str}) becomes
\begin{equation}
       \big[1+\frac{1}{2}M\big(D_{x}, 0\big)\big] v_{t}+v_{x}+\mu v^{2}v_{x}=0. \label{sBKP-uni}
\end{equation}
 So, any $y$-independent solutions of (\ref{sBKP-str}) are  also solutions of (\ref{sBKP-uni}).

 As in the preceding section, let us consider the simplest case $M\big(D_{x}, D_{y}\big)=-2 \nu D_x^2$ for which    (\ref{sBKP-str}) reduces to
\begin{equation}
        \Big(v_{t}+v_{x}+ \mu v^{2}v_{x}-\nu v_{xxt}\Big)_{x}+\frac{1}{2}v_{yy}=0. \label{B-ins}
\end{equation}
With a quadratic nonlinearity but not with a cubic one, this equation has been proposed as a model equation in different contexts (for instance see \cite{Min2007} for water waves). It is called the BBM-KP equation or the regularized version of the KP equation \cite{Bona2002}. In the remainder of this subsection we study the linear transverse instability of the unidirectional solitary wave solutions of (\ref{B-ins}).

If we neglect transverse effects, (\ref{B-ins}) reduces to the cubic BBM equation
\begin{equation}
        v_{t}+v_{x}+ \mu v^{2}v_{x}-\nu v_{xxt}=0. \label{bbm3}
\end{equation}
Consider a travelling wave solution of (\ref{bbm3}) of the form $v(x,t)=\alpha R(\xi)$, $\xi = x - ct$, where $\alpha$ and $c$ are constants. Plugging this into (\ref{bbm3}), we see that $R$  satisfies the ordinary differential equation given by (\ref{Nop}) with $\mu>0$, $\nu >0$, $c-1>0$ and
\begin{equation}
       \alpha^2=6(c-1)/\mu,~~~~\kappa^{2}=(c-1)/(\nu c). \label{bbm-sol}
\end{equation}
So the cubic BBM equation  has solitary wave solutions in the form $v(x,t)=\alpha R(x-ct)=\alpha \sech \big(\kappa (x-ct)\big)$. Notice that the only difference between the solitary wave solutions corresponding to (\ref{bbm3}) and (\ref{mkdv}) is due to the definition of the parameter $\kappa$. We now  repeat the linear instability analysis in a fashion very similar to that in Section \ref{sec4}. So we consider small transverse perturbations of the unidirectional solitary wave solution and assume that the perturbations are periodic in $y$ with long wavelength. If we substitute (\ref{pertur})  into (\ref{B-ins}) (keeping the  definition  given by (\ref{bbm-sol}) for $\kappa$  in mind) and linearize the resulting equation, we  get  the fourth-order linear  ordinary differential equation
\begin{equation}
         \frac{d^2}{d \xi^2}{\cal L} z=\frac{ \Omega}{c} \frac{d^3z}{d \xi^3}
  -\frac{\Omega}{\nu c} \frac{d z}{d \xi}+   \frac{\lambda^2}{2\nu c} z, \label{odebbm}
\end{equation}
where the linear operator ${\cal L}$ is given in (\ref{oper}),  $\xi=x-ct$ is  the travelling wave coordinate, $\lambda$ is the transverse wave number and $\Omega$ is possibly a complex constant.  We now repeat the argument in the previous section. We again assume that $z$ and its derivatives vanish in the limit $\xi \rightarrow \pm \infty$. As in the previous section, we are interested in long-wavelength transverse perturbations. So, under the assumption of  $\vert \lambda \vert <<1$, we again assume the small parameter expansion given in (\ref{expan}). At the leading order, substitution of  the series expansions (\ref{expan})   into (\ref{odebbm})   leads to the homogeneous equation (\ref{zeroth}). A decaying solution of (\ref{zeroth}) is again $z_0(\xi)=a_0 (\sech (\kappa\xi))^{\prime}=a_0 R^{\prime}(\xi)$ where $a_{0}$ is an arbitrary constant. At ${\cal O}(\lambda)$ we get the following inhomogeneous differential equation for $z_{1}$
\begin{equation}
   \frac{d^2}{d \xi^2}\mathcal{L} z_1       =\frac{ \Omega_{1}}{c} \frac{d^3z_{0}}{d \xi^3}
  -\frac{\Omega_{1}}{\nu c} \frac{d z_{0}}{d \xi}. \label{bbma}
 \end{equation}
 If we substitute $z_0(\xi)=a_0 R^{\prime}(\xi)$ to the right-hand side, integrate twice and use the zero conditions at infinity, we obtain
 \begin{equation}
        \mathcal{L}z_1=\frac{\Omega_{1}a_{0}}{ c} R^{\prime\prime}(\xi)-\frac{\Omega_{1}a_{0}}{\nu c} R(\xi). \label{firstb}
 \end{equation}
 A decaying solution of this inhomogeneous differential equation is
 \begin{equation}
    z_1(\xi)= a_1R^{\prime}(\xi)+\frac{\Omega_{1}a_{0}}{2\kappa^{2}\nu c}(\kappa^2\nu-1)\xi R^{\prime}(\xi)+ \frac{\Omega_{1}a_{0}}{2\kappa^{3}\nu c }R^{\prime}(\xi)\coth (\kappa\xi), \label{firsts}
 \end{equation}
 where $a_{1}$ is an arbitrary constant. At ${\cal O}(\lambda^{2})$ we get the differential equation
 \begin{equation}
        \frac{d^2}{d \xi^2}\mathcal{L} z_2
                        = \frac{1}{c} \frac{d^{3}}{d \xi^3}\big(\Omega_{2}z_0             +\Omega_{1}z_1\big)
                        -\frac{1}{\nu c} \frac{d}{d \xi}\big(\Omega_{2} z_0
                        +\Omega_{1} z_1\big)
                        +\frac{1}{2\nu c} z_0 \label{secondBBM}
 \end{equation}
 for $z_{2}$. We now derive a solvability condition  by multiplying both sides of (\ref{secondBBM}) by $r(\xi)$ with $r^{\prime}=R$  and then by integrating over the entire real line. From the left-hand side of  (\ref{secondBBM}) we again have $\big\langle (\mathcal{L}z_2)^{\prime\prime},r\big\rangle =0$  as in (\ref{left}).  The values of the integrals associated with the terms on the right-hand side of (\ref{secondBBM}) are calculated as follows:
 \begin{eqnarray*}
       &&  \!\!\!\!\!\!\!\!\!\!\!
       \big\langle z_{0},r\big\rangle = -\frac{2}{\kappa}a_{0}, ~~~~~~
       \big\langle z_{0}^{\prime},r\big\rangle  = 0, ~~~~~~
       \big\langle z_{0}^{\prime\prime\prime},r\big\rangle =0\\
      && \!\!\!\!\!\!\!\!\!\!\!
      \big\langle z_{1}^{\prime},r\big\rangle= \frac{\Omega_{1}a_{0}}{2\kappa^{3}\nu c}(\kappa^{2}\nu+1),~~~~~~
      \big\langle z_{1}^{\prime\prime\prime},r\big\rangle
                                =\frac{\Omega_{1}a_{0}}{6\kappa \nu c}(\kappa^{2}\nu-3).
 \end{eqnarray*}
 Combining the above results  we get the solvability condition for (\ref{secondBBM}) as
 \begin{equation}
                0=\frac{\Omega_{1}^{2}a_{0}}{6\kappa \nu c^{2}}(\kappa^{2}\nu-3)
                -\frac{\Omega_{1}^{2}a_{0}}{2\kappa^{3}\nu^{2} c^{2}}(\kappa^{2}\nu+1)
                - \frac{a_{0}}{\kappa\nu c}.  \label{solvb1}
 \end{equation}
 or, with $a_{0}\neq 0$
 \begin{equation}
               \Omega_{1}^{2}=-\frac{6c^{2}(c-1)}{4(2c+1)(c-1)+3}<0,     \label{solvbf}
 \end{equation}
 where we have used $\kappa^{2}=(c-1)/(\nu c)$. The result $\Omega_{1}^{2}<0$ for any $c>1$ means that at this order of approximation our analysis does not demonstrate the existence of linear transverse instability of the  one-dimensional solitary wave solutions of (\ref{B-ins}) under the long wavelength periodic transverse perturbations. This is in contrast to the situation observed  in the previous section for the simplified form of the Whitham-type KP equation.  It is worth mentioning here that linear transverse instability of the line solitary waves  of (\ref{B-ins}) may arise  at  higher-order approximations.  We also remind that the present  analysis is  restricted to considering long-wavelength transverse perturbations. It remains an open question whether the assumption that the transverse perturbations have long wavelength  can be removed and  whether this will change the result obtained above.

\bibliographystyle{plainnat}
\bibliography{WM-revised-references}

\end{document}